\documentclass[aps,twocolumn,superscriptaddress]{revtex4}
\usepackage{graphicx}

\newcommand{\bra}[1] {\left\langle #1 \right|}
\newcommand{\ket}[1] {\left| #1 \right\rangle}
\newcommand{\prodket[2]} {\left| #1 \right\rangle \otimes
\left| #2 \right\rangle}

\begin{document}

\title{Observing quantum non-locality in the entanglement between
modes of massive particles}


\

\author{S. Ashhab}
\affiliation{Frontier Research System, The Institute of Physical
and Chemical Research (RIKEN), Wako-shi, Saitama 351-0198, Japan}

\author{Koji Maruyama}
\affiliation{Frontier Research System, The Institute of Physical
and Chemical Research (RIKEN), Wako-shi, Saitama 351-0198, Japan}
\affiliation{Laboratoire d'Information Quantique and QUIC, CP
165/59, Universit\'{e} Libre de Bruxelles, 1050 Bruxelles,
Belgium}

\author{Franco Nori}
\affiliation{Frontier Research System, The Institute of Physical
and Chemical Research (RIKEN), Wako-shi, Saitama 351-0198, Japan}
\affiliation{Physics Department and Michigan Center for
Theoretical Physics, The University of Michigan, Ann Arbor,
Michigan 48109-1040, USA}

\date{\today}

\begin{abstract}
We consider the question of whether it is possible to use the
entanglement between spatially separated modes of massive
particles to observe nonlocal quantum correlations. Mode
entanglement can be obtained using a single particle, indicating
that it requires careful consideration before concluding whether
experimental observation, e.g. violation of Bell inequalities, is
possible or not. In the simplest setups analogous to optics
experiments, that observation is prohibited by fundamental
conservation laws. However, we show that using auxiliary
particles, mode entanglement can be converted into forms that
allow the observation of quantum non-locality. The probability of
successful conversion depends on the nature and number of
auxiliary particles used. In particular, we find that an auxiliary
Bose-Einstein condensate allows the conversion arbitrarily many
times with a small error that depends only on the initial state of
the condensate.
\end{abstract}


\maketitle

\section{Introduction}

Entanglement has attracted interest ever since it was recognized
as a central ingredient in quantum mechanics \cite{Schroedinger},
both on a fundamental level as it represents a clear deviation
from classical intuition \cite{Einstein} and, recently, for its
possible use as a resource in the emerging field of quantum
information processing \cite{QIP_Reviews}.

One typically thinks of entanglement as describing quantum
correlations between physical variables of two distinct particles
or systems. Perhaps the simplest example is the Bell states of two
(distinguishable) spin-1/2 particles. There are, however, a number
of other situations where entanglement arises. For example, in the
Stern-Gerlach experiment, the position of a particle is entangled
with its spin. Another example, which will be the main focus of
the present paper and will be explained shortly, is the so-called
mode entanglement. On a more abstract level, one could even
redefine a single physical degree of freedom (rather unnaturally)
such that certain superpositions of basis states have the
appearance of entangled states \cite{Hirayama}.

Mode entanglement can be perhaps most easily understood as
follows: let us take a single particle in a quantum superposition
of two basis states. In the field-theory description, we treat the
modes (or, in other words, the basis states) as the quantum
objects: each mode can be in the state of being occupied by zero
particles, one particle and so on. Particles are therefore seen as
excitations of the field modes. Going back to our example with a
single particle in a superposition state, we find the alternative
description where the two modes are in an entangled state: a
quantum superposition where one mode is occupied and the other
mode is empty, with the two possible combinations present in the
superposition.

Given the wide range of manifestations of entanglement explained
above, not all of them are equally intriguing. It is in fact the
combination of entanglement and non-locality that fascinates us
physicists the most. In particular, the entanglement between two
degrees of freedom of a single object can be easily overlooked in
this context because it does not contradict intuitive expectations
about locality. One must therefore divide entanglement into two
types: loosely speaking, one can speak of interesting versus
uninteresting entanglement.

Clearly the above classification of the different types of
entanglement according to how interesting they are is not
unambiguous. One must therefore define that classification
according to an unambiguous experimental procedure. It seems to us
that the most reasonable definition would include the observation
of nonlocal, non-classical correlations between two objects. The
term `nonlocal' is used in this context to mean the following: if
we are to say that there is nonlocal entanglement (in an
unquestionable form) between two objects at a given point in time,
we require that quantum correlations be observable within a time
$d/c$, where $d$ is the distance between the two objects, and $c$
is the speed of light \cite{Einstein,Bell,NonlocalityWarning}.

An important requirement needed to detect entanglement is the
ability to perform measurements in different bases. For example,
if we take an ensemble of pairs of spin-1/2 particles and we find
that the ensemble has perfect correlations (within each pair) in
the measured values of the spins along the $z$-axis, we still
cannot say for sure that the pairs are entangled. These
observations would be consistent with classical correlations. Only
after performing measurements along several different pairs of
directions can we establish the presence of quantum correlations.
The requirement of measurements in different bases is in fact the
difficulty when dealing with modes of massive particles. We can
measure the number of particles in a mode, but we cannot measure
superpositions of particle numbers, as will be explained below
\cite{Wick,Vestraete,Measuring_modes}.

In this paper, we focus on the case of entanglement between
spatially separated modes of massive particles. Since, as we have
just mentioned, direct measurements on the modes cannot be used to
observe nonlocal quantum effects, we ask the question of whether
this type of entanglement can be converted into a different type
such that nonlocal quantum effects are observable. In spite of the
fundamental difficulties associated with particle conservation,
some recent studies \cite{TerraCunha} have suggested that this
conversion is in fact possible. Here we analyze several possible
methods to {\it observe nonlocal quantum correlations between
spatially separated modes without violating any fundamental
superselection rules}. In particular, we find the rather
surprising result that a finite auxiliary Bose-Einstein condensate
(BEC) can be used to convert this single-particle entanglement an
arbitrary number of times, neglecting difficulties that could
arise in a realistic setup.

In fact, this type of single-particle entanglement has attracted
significant interest in recent years
\cite{Tan,Hardy,Wiseman,Hessmo,Babichev,vanEnk,Bartlett,DAngelo}.
There seems to be a consensus over the entanglement in a single
photon, as will be explained briefly below. However, our proposals
presented in this paper achieve the conversion of entanglement in
a single massive particle. Our proposals can also be seen as
concrete realizations of the recently proposed concept of
bound-entanglement activation \cite{Horodecki}.

The role of the BEC in this problem can be understood, rather
figuratively, as follows: the presence of the single (delocalized,
or flying) particle can be `captured on a quantum film', i.e.
recorded in the state of a quantum object, such that no collapse
of the wave function occurs during this process. One `quantum
film' is placed on each side of the apparatus, so that one of them
will record the presence of the flying particle. However, quantum
correlations cannot be observed between these `quantum films',
because the particle still carries the which-path information
about its location. The BEC can now be seen, in some sense, as a
sea of particles identical to the one carrying the which-path
information. If the BEC is prepared in a suitable initial state
and the flying particle is properly injected into this sea of
identical particles, the which-path information is lost (i.e.
truly erased), and quantum correlations can be observed when we
measure (or, if you wish, develop) the `quantum films'. The
perfect sea of particles for discarding the flying particle would
be a coherent state with many particles. One should note, however,
that true coherent states (without a well-defined total particle
number) do not exist in systems of massive particles; coherent
states usually provide a useful calculational tool, but they
should be used with care when dealing with conceptual questions
\cite{LeggettNote,Kuklov}. This is the reason why one must
construct the initial state of the BEC such that it emulates, as
much as possible, a coherent state for the purposes of the
proposed experiment, while still obeying particle-number
conservation laws.

The paper is organized as follows: In Sec. II we present a simple
experimental setup where the fundamental questions about the
possibility of detecting the entanglement between modes can be
seen clearly. In Sec. III we present a quantum-eraser approach to
detecting the entanglement, and we discuss why we are not
interested in that approach. In Sec. IV we discuss how the
entanglement can be converted using auxiliary atoms. Section V
contains concluding remarks.

\section{Basic problem}

Take a particle-beam-splitter setup, as shown in Fig. 1(a) for the
case of a photon beam splitter. A particle passes through the beam
splitter, resulting in the quantum state:

\noindent
\begin{equation}
\ket{\Psi} = \frac{1}{\sqrt{2}} \left( \ket{L} + \ket{R} \right).
\label{eq:Psi_after_BS}
\end{equation}

\begin{figure}[h]
\vspace{0.0cm}
\includegraphics[width=7.0cm]{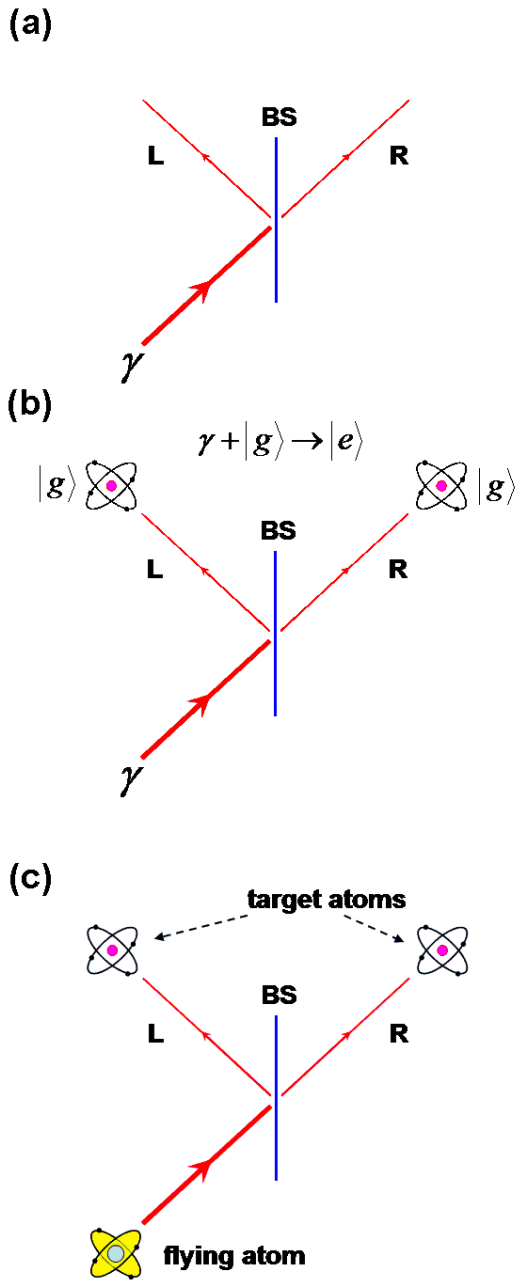}
\caption{(color online) Schematic diagrams of beam-splitter
setups. Figure 1(a) shows the basic optics setup. Figure 1(b)
shows a photon beam-splitter setup with two target atoms, one of
which will be excited by absorbing the incident photon. Figure
1(c) shows the atomic analog of Fig 1(b), where the flying atom
excites one of the two target atoms.}
\end{figure}

\noindent When viewed as a quantum state of the particle, the
above state clearly does not seem to contain any entanglement. If
one thinks of the modes of the field describing that species of
particles, however, one finds that the modes on the left and right
sides of the beam splitter are entangled; the state is a quantum
superposition of the mode on the left containing one particle and
the mode on the right empty as well as the opposite combination.
The question now is whether this type of entanglement is simply a
book-keeping issue that has no physical implications or it leads
to measurable effects associated with entanglement, e.g. violation
of the Bell inequalities using local measurements \cite{Bell}. If
the particle is a photon, one immediately finds that by
positioning two atoms in their ground states, one on each side of
the beam splitter, it is possible to design the system such that
the photon will excite the corresponding atom. This process is
illustrated in Fig. 1(b), and it results in the quantum state

\noindent
\begin{equation}
\ket{\Psi} = \frac{1}{\sqrt{2}} \left( \ket{eg} + \ket{ge}
\right), \label{eq:Psi_after_absorption}
\end{equation}

\noindent where the symbols $g$ and $e$ stand for ground and
excited state, respectively, and the first and second symbols
describe the atoms on the left and right, respectively. It is now
possible to measure the state of each atom in any desired basis
(for example, a combination of unitary operations using classical
electromagnetic fields and a measurement in the
$\{\ket{g},\ket{e}\}$ basis will do). One should therefore be able
to observe a violation of the Bell inequalities when measuring the
states of the two atoms \cite{MeasurementNote}.

We now ask whether a similar procedure can be followed in the case
of massive particles. A key point to note here is that nature
allows us to create and destroy photons. In other words, there is
no conservation law for the number of photons, and there are
physical processes that change it, e.g. absorption of a photon by
an atom. However, there are conservation laws for all massive
particles (here we shall not get into high-energy-physics
discussions, and we shall pretend, for simplicity, that there is a
field describing whatever particle we are considering, e.g.
atoms).

We now imagine the situation where a flying atom passes through
the beam splitter \cite{Experiment} and ends up in the quantum
state of Eq. (\ref{eq:Psi_after_BS}). We take two target atoms as
above, and we design the apparatus such that the flying atom
interacts with and excites the corresponding target atom, as shown
in Fig. 1(c). Instead of the state in Eq.
(\ref{eq:Psi_after_absorption}), one now finds the state

\noindent
\begin{equation}
\ket{\Psi} = \frac{1}{\sqrt{2}} \left( \prodket[L]{eg} +
\prodket[R]{ge} \right). \label{eq:Psi_after_excitation}
\end{equation}

\noindent In order to predict the correlations that would be
observed in any measurement on the target atoms, one must trace
out the flying atom's degree of freedom. One therefore finds the
reduced density matrix

\noindent
\begin{equation}
\rho = \left(
\begin{array}{cccc}
0 & 0 & 0 & 0 \\
0 & \frac{1}{2} & 0 & 0 \\
0 & 0 & \frac{1}{2} & 0 \\
0 & 0 & 0 & 0
\end{array}
\right),
\end{equation}

\noindent and the measurement results will be classically
correlated. No signature of entanglement can be observed in these
results.

From the above example, it is clear that tracing out the degree of
freedom of the flying atom prevents us from observing quantum
correlations between the target atoms. We can therefore say that
the entanglement between the modes in the above setup cannot be
tested. In the following sections, we shall try to construct more
elaborate setups such that entanglement between the modes of the
flying atom can result in measurable entanglement between the
target atoms. In the first case, we will use a global measurement
on the state of the flying atom. In the other two cases, auxiliary
modes of the same species as the flying atom will be used.

\section{Quantum-eraser approach}

We start with perhaps the simplest case. As we discussed above,
tracing out the degree of freedom of the flying atom results in
classical correlations between the states of the target atoms, but
no quantum correlations would be observable. Performing a
measurement on the flying atom while it is on one side of the
apparatus does not help either, since such a measurement would
project the state of the target atoms onto one of two separable
state. However, if we perform a measurement on the flying atom
that erases the information about which side of the beam splitter
it went into, the target atoms would end up in an entangled state.
The details are given below.

Let us take the situation where the flying atom was prepared in
the superposition state given by Eq.~(\ref{eq:Psi_after_BS}), and
then excited the corresponding target atom. We therefore have the
quantum state of Eq.~(\ref{eq:Psi_after_excitation}). We now
perform a measurement on the flying atom in the basis $\left\{
 \ket{+} \equiv (\ket{L}+\ket{R})/\sqrt{2}, \;\, \ket{-} \equiv
(\ket{L}-\ket{R})/\sqrt{2} \right\}$. Depending on the outcome of
the measurement, the target atoms end up in the state

\noindent
\begin{equation}
\ket{\Psi} = \frac{1}{\sqrt{2}} \left( \ket{eg} \pm \ket{ge}
\right),
\end{equation}

\noindent where the upper and lower signs correspond to the
outcomes $\ket{+}$ and $\ket{-}$, respectively. By post-selecting
only those instances where the outcome $\ket{+}$ was obtained,
quantum correlations would be observable between the target atoms.
Naturally one could equally well post-select the instances where
the outcome $\ket{-}$ was obtained.

Although the above procedure allows us, in a sense, to observe the
entanglement between the modes of the flying atom, it does not
meet the criteria we set in Sec. I above: In order to observe the
entanglement, we had to perform a global measurement on the state
of the flying atom. In other words, after the flying atom excited
one of the target atoms, it had to travel (carrying quantum
information) to a common detection location, so that its state
could be measured in the desired basis. One could therefore argue
that the entanglement did not exist in any meaningful form at the
time of excitation, but rather the flying atom and measurement
apparatus mediated an interaction between the target atoms.
Accounting for the time required to establish the measurement
results then violates the criteria of Sec. I. We shall therefore
not allow such global measurements in the following sections.

\section{Observing entanglement without global measurements}

As above, we take a flying atom going through a beam splitter and
exciting one of two target atoms. Since the flying atom cannot be
annihilated by any physical process, this simple setup cannot be
used to detect entanglement between the target atoms. We now
present two scenarios where auxiliary atoms of the same species as
the flying atom can be used to erase, at least partially, the
information about which side of the beam splitter the flying atom
went into.

\subsection{Scenario I: one auxiliary atom}

Let us imagine that we have already prepared an atom of the same
species as the flying atom in the state

\noindent
\begin{equation}
\ket{\Psi_{\rm aux}} = \frac{1}{\sqrt{2}} \left( \ket{L_{\rm aux}}
+ \ket{R_{\rm aux}} \right).
\end{equation}

\noindent where the states $\ket{L_{\rm aux}}$ and $\ket{R_{\rm
aux}}$ describe localized modes on the left and right side of the
beam splitter, respectively. The quantum state of the entire
system can therefore be expressed as

\noindent
\begin{eqnarray}
\ket{\Psi} & = & \frac{1}{2} \Big( \ket{L_{\rm aux}} + \ket{R_{\rm
aux}} \Big) \nonumber \\ & & \otimes \Big( \prodket[L_{\rm
flying}]{eg} + \prodket[R_{\rm flying}]{ge} \Big).
\label{eq:Psi_after_excitation_aux}
\end{eqnarray}

\noindent We now wish to use the indistinguishability between the
flying atom and the auxiliary atom to erase (even if partially)
the which-path information carried by the flying atom. We find it
instructive to start with what we consider a non-ideal approach,
especially for those who might not be sufficiently familiar with
certain details of the procedures discussed below.

\subsubsection{Non-ideal approach}

Let us imagine that the flying atom is stopped and trapped after
it has excited one of the target atoms (there is no fundamental
difficulty with doing that). The flying and auxiliary atoms are
now trapped in the ground states of localized potential wells. If
we take the quantum state in Eq.
(\ref{eq:Psi_after_excitation_aux}) and slowly merge the two wells
on the left and the two wells on the right, one might expect the
resulting state to be given by

\noindent
\begin{equation}
\ket{\Psi} = \frac{1}{2} \Big( \prodket[2,0]{eg} + \ket{1,1}
\otimes \left\{ \ket{eg} + \ket{ge} \right\} + \prodket[0,2]{ge}
\Big), \label{eq:Psi_after_excitation_aux_merged}
\end{equation}

\noindent where the ket on the left describes the total number of
itinerant atoms (i.e. flying and auxiliary) on the left and right
sides of the beam splitter. Assuming that the above is true, if we
measure the number of atoms in the combined traps, we would have a
50\% chance of finding one atom on each side of the beam splitter.
If that happens, we would have lost all information about which
side the flying atom went into. We would therefore find the state
of the target atoms to be

\noindent
\begin{equation}
\ket{\Psi} = \frac{1}{\sqrt{2}} \left( \ket{eg} + \ket{ge}
\right),
\end{equation}

\noindent and quantum correlations would be observable in those
instances. In the instances where the two atoms (flying and
auxiliary) are found on the same side of the beam splitter, we
know which side the flying atom went into, and we end up with a
separable state for the target atoms.

\begin{figure}[h]
\vspace{0.0cm}
\includegraphics[width=7.0cm]{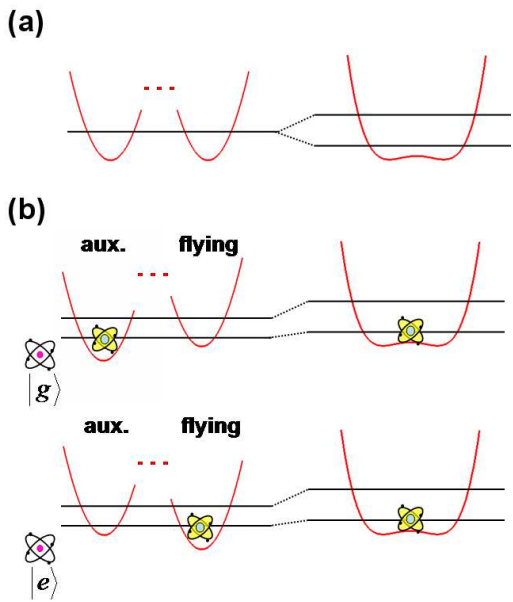}
\caption{(color online) Merging two wells into one (note that the
two wells being merged are on the same side of the beam splitter).
If the initial double-well potential is perfectly symmetric with
an infinite barrier (a), the ground state is degenerate, and it
splits into two states during the merging process. The symmetry
can be lifted using internal-state-dependent interactions with the
target atom, ensuring that the atom ends up in the ground state of
the merged well (b).}
\end{figure}

Although the above scenario might, at first sight, look like it
does not violate any principles of quantum mechanics, closer
inspection reveals the opposite. In particular, let us take the
situation where one atom is on each side of the beam splitter
before merging the two pairs of wells. It would be wrong to
conclude that the slow merging process results in the state
$\ket{1,1}$ (without any additional quantum numbers). If we note
that two orthogonal states, namely $\ket{L_{\rm aux.}} \otimes
\ket{R_{\rm flying}}$ and $\ket{R_{\rm aux.}} \otimes \ket{L_{\rm
flying}}$, would result in the same state according to that
description, we find that the process of merging the wells must be
treated with more care. Let us focus on a pair of wells with one
atom in the ground state of one of them. If we take the limit of
an infinite barrier height (and zero bias between the two wells),
the ground state of the double-well potential is degenerate (see
Fig. 2(a)). When we combine the two wells, the atom will end up in
a superposition between the ground and first-excited states of the
(merged) single well, with amplitudes and a phase factor that
depend on the source of the atom (i.e., flying or auxiliary atom)
and the details of the merging process. The which-path information
is therefore not erased at this stage. It is possible, in
principle, to perform certain measurements on the atoms in the
merged wells and establish the phase in the entangled state of the
target atoms (similarly to what was done in the quantum-eraser
approach above). However, since (1) post-selection involving the
exchange of information between the two sides would be required in
this case and (2) the state resulting from the merging of the
wells is strongly susceptible to fluctuations in the bias between
the two separate wells, we abandon this approach.

Note that since we only obtain the desired entangled state when
the flying atom and auxiliary atom are found on opposite sides of
the beam splitter, the difference between bosonic and fermionic
particles does not affect our argument. It only affects the
instances where both atoms are found on the same side of the beam
splitter. But those instances will be discarded. We therefore do
not need to analyze the two cases separately.

\subsubsection{Better approach}
\label{Sec:Correct_approach}

Let us now take the same procedure as in the `non-ideal approach'
above, except that during the well-merging process we bring the
target atoms close to the merging wells. We design the system such
that a target atom in the ground state raises the well of the
flying atom relative to that of the auxiliary atom, and a target
atom in the excited state lowers the well of the flying atom
relative to that of the auxiliary atom, as shown in Fig. 2(b).
Following this procedure each itinerant atom (assuming one atom
was on each side of the beam splitter) will be in the ground state
of its potential well. The quantum state is then correctly
described by Eq. (\ref{eq:Psi_after_excitation_aux_merged}); the
desired (partial) erasure of the which-path information has been
achieved. We therefore obtain an entangled pair of target atoms
50\% of the time and separable states 50\% of the time. The
desired raising and lowering of the potential wells could be
achieved using electronic-state-dependent interactions between the
itinerant and target atoms.

Three remarks are in order: (1) When merging the two wells, one
must avoid back action from the atom on the trapping apparatus.
That back action would in effect perform a measurement on the
location of the flying atom, preventing the observation of
entanglement. Since the trapping apparatus is macroscopic, its
recoil from moving the trapped atom should be negligible if
designed properly (in particular, one must also avoid resonant
scattering in the case of an optical atomic trap). (2) We still
discard 50\% of the samples, namely when we find both itinerant
atoms on the same side. However, this step does not raise the same
concerns that we raised regarding post-selection above, because
the decision of discarding a given sample can be made locally,
without any exchange of information between the two sides. (3) The
two-well-merging process is essentially equivalent to a standard
controlled-NOT gate on two qubits. Before the process, the target
atom and the position of the trapped (i.e. flying or auxiliary)
atom in the double well are perfectly correlated. Depending on the
state of the target atom, i.e. the control bit, we change the
position of the trapped atom, such that the final state of the
trapped atom is disentangled from that of the target atom. This is
exactly the same information erasing process as what was described
in Ref. \cite{Maruyama} in the context of Maxwell's demon.

\subsubsection{Alternative approach: initially correlated flying and auxiliary atoms}

For completeness let us mention the following possibility. If we
take the auxiliary atom and the flying atom to go through the beam
splitter immediately after each other such that their interaction
ensures that each one of them goes to a different side of the beam
splitter, we are guaranteed to have the favourable situation with
one itinerant atom on each side of the beam splitter. We are thus
guaranteed to obtain entanglement between the target atoms.
However, the entanglement after crossing the beam splitter in this
situation is closer to being an entanglement involving the
internal degrees of freedom of two particles than it is to the
entanglement between modes that we are interested in. In this
case, the `internal' degree of freedom is the time at which the
atom goes through the beam splitter: if the atom on the left
crossed the beam splitter at the earlier time, the atom on the
right must have crossed the beam splitter at the later time, and
vice versa. We therefore do not consider this possibility in any
more detail.

\subsection{Scenario II: Auxiliary Bose-Einstein condensate}

The above scenario allows us to obtain an entangled pair of target
atoms 50\% of the time. Since only one auxiliary atom was used
there, it is natural to ask whether replacing the auxiliary atom
with an auxiliary BEC would result in a higher probability for
producing a pair with the desired entanglement. Furthermore, one
could still argue that when using an auxiliary atom one is not
probing the entanglement from the flying atom, but rather a
complicated form of entanglement involving both the flying and
auxiliary atom. These issues will be addressed below.

Before proceeding let us also note here that the conversion, or
transfer, of entanglement from the flying atom to the target atoms
was made possible by the fact that the auxiliary atom was not
localized in the initial state. The quantum fluctuations
associated with our lack of knowledge about the number of
auxiliary atoms on the left and right sides of the beam splitter
allowed us to erase the which-path information carried by the
flying atom. After the merging of the wells (and discarding the
instances where both flying and auxiliary atoms turned out on the
same side of the beam splitter), there is no way, even in
principle, to tell whether the flying atom went to the left or
right. A natural generalization would therefore be to use a large
number of auxiliary atoms where the addition of a single atom
would be hardly noticeable. One is therefore led to think of
coherent states. As mentioned in Sec. I, however, truly coherent
states cannot be prepared in this system. The closest one can get
to them is a single BEC (with a well-defined number of atoms)
shared between the two sides of the beam splitter, such that it is
allowed to have quantum fluctuations in the number of atoms on
each side.

We now present a procedure based on the above-mentioned idea. We
take a BEC of non-interacting atoms trapped in a single well, and
we split the well into two and take those to opposite sides of the
beam splitter. The state of the BEC is given by

\noindent
\begin{eqnarray}
\ket{\Psi_{\rm BEC}} & = & \frac{1}{\sqrt{N!}} \left(
\frac{a_{L,\rm aux}^{\dagger} + a_{R,\rm aux}^{\dagger}}{\sqrt{2}}
\right)^N \ket{\rm vac} \nonumber
\\
& = & \sum_{j=0}^{N} \sqrt{f_{N,j}} \ket{j,N-j}_{\rm aux}
\label{eq:BEC_initial_state}
\end{eqnarray}

\noindent where the operators $a_{L,\rm aux}^{\dagger}$ and
$a_{R,\rm aux}^{\dagger}$ create atoms of the same species as the
flying atom in auxiliary modes on the left and right sides of the
beam splitter, respectively, $N$ is the total number of atoms in
the condensate, and $\ket{\rm vac}$ is the vacuum state for that
species \cite{Andrews,Leggett,Hines}. The function $f_{N,j}$ is
defined as

\noindent
\begin{equation}
f_{N,j} \equiv \frac{1}{2^N} \times \frac{N!}{j! (N-j)!}.
\end{equation}

\noindent In the ket of the second line of Eq.
(\ref{eq:BEC_initial_state}), the first and second numbers denote
the number of atoms in the BEC on the left and right sides of the
beam splitter, respectively (note that the total number of atoms
in the entire system is well defined, in accordance with
superselection rules; the number of atoms in the left and right
wells, however, is not well defined). The state of the entire
system is therefore given by

\noindent
\begin{eqnarray}
\ket{\Psi} & = & \sum_{j=0}^{N} \sqrt{f_{N,j}} \ket{j,N-j}_{\rm
aux} \nonumber \\ & & \otimes \frac{1}{\sqrt{2}} \left(
\prodket[L_{\rm flying}]{eg} + \prodket[R_{\rm flying}]{ge}
\right).
\end{eqnarray}

\noindent We now follow a similar approach to the auxiliary-atom
scenario of Sec. \ref{Sec:Correct_approach} and slowly merge the
trapping well of the BEC with that of the flying atom
\cite{Merging_BEC}. We can now express the state of the system as

\noindent
\begin{eqnarray}
\ket{\Psi} & = & \frac{1}{\sqrt{2}} \Bigg(
\frac{1}{2^{N/2}} \prodket[N+1,0]{eg} + \frac{1}{2^{N/2}}
\prodket[0,N+1]{ge} \nonumber \\
& & \hspace{0.5cm} + \sum_{j=1}^{N} \ket{j,N+1-j} \nonumber \\ & &
\hspace{1cm} \otimes \left\{ \sqrt{f_{N,j-1}} \ket{eg} +
\sqrt{f_{N,j}} \ket{ge} \right\} \Bigg).
\label{eq:Psi_after_excitation_BEC_merged}
\end{eqnarray}

\noindent By tracing over the BEC degree of freedom, we find that
the relevant off-diagonal element of the target atoms' reduced
density matrix is given by

\noindent
\begin{eqnarray}
\rho_{ge,eg} & = & \frac{1}{2} \sum_{j=1}^{N}
\sqrt{f_{N,j-1}} \times \sqrt{f_{N,j}} \nonumber \\
& = & \frac{1}{2} \sum_{\delta j=-(\frac{N}{2}-1)}^{\frac{N}{2}}
f_{N,\frac{N}{2}+\delta j} \times \sqrt{\frac{\frac{N}{2}
+ \delta j}{\frac{N}{2}- \delta j+1}} \nonumber \\
& \approx & \frac{1}{2} \times \left(1 - \frac{1}{2N} \right),
\end{eqnarray}

\noindent where we have defined $\delta j \equiv j-N/2$ and
assumed that $N$ is even in the intermediate steps. Combining the
above expression for $\rho_{ge,eg}$ with the fact that
$\rho_{ge,ge}=\rho_{eg,eg}=1/2$, the probability of successfully
generating the desired entangled state is found to approach unity,
with an error that decreases as $1/N$. To put it differently, the
concurrence $C$, which can be calculated straightforwardly
\cite{Wootters}, is given by

\noindent
\begin{equation}
C = 1 - \frac{1}{2N}.
\end{equation}

\noindent The above expression for the concurrence turns out to be
a good approximation even for small values of $N$, with accidental
agreement for $N=1$.

It would seem wasteful to use a large condensate to generate a
single pair of entangled target atoms. The next question to ask is
therefore how many times we can use the same condensate to
generate entangled pairs of target atoms from a stream of flying
atoms. In order to answer this question, we take the quantum state
in Eq. (\ref{eq:Psi_after_excitation_BEC_merged}) and trace out
the target atoms' degrees of freedom. We find that the BEC reduced
density matrix is given by

\noindent
\begin{equation}
\rho_{\rm BEC}^{\rm after \ 1st \ run} = \frac{1}{2} \left(
\ket{\Lambda_L} \bra{\Lambda_L} + \ket{\Lambda_R} \bra{\Lambda_R}
\right), \label{eq:reduced_density_matrix}
\end{equation}

\noindent where

\noindent
\begin{eqnarray}
\ket{\Lambda_L} & = & \sum_{j=0}^{N} \sqrt{f_{N,j}} \ket{j+1,N-j}
\nonumber
\\
\ket{\Lambda_R} & = & \sum_{j=0}^{N} \sqrt{f_{N,j}} \ket{j,N+1-j}.
\end{eqnarray}

\noindent The subscript of $\Lambda$, i.e. L or R, indicates the
path that the first flying atom took, which is also copied to the
state of the target atoms. The density matrix in Eq.
(\ref{eq:reduced_density_matrix}) describes a mixed state, which
might suggest that the probability of successful production of a
second entangled pair of target atoms will be reduced. We note,
however, that each of the two pure states $\ket{\Lambda_L}$ and
$\ket{\Lambda_R}$ has exactly the same atom-number distribution as
the original BEC state, except that the center of the distribution
can be shifted. If we follow the same procedure as above with the
BEC in this new mixed state, the result can be obtained by taking
the average of the expected results from states $\ket{\Lambda_L}$
and $\ket{\Lambda_R}$. Since each one of those gives identical
results to the case of a pure BEC state with $N$ atoms, we find
that the second flying atom produces an entangled state between
the target atoms with the same error (of order $1/N$) as the first
flying atom. The procedure can be repeated indefinitely, with the
error remaining at $1/(2N)$, where $N$ is the initial number of
atoms in the condensate. It is quite remarkable that we obtain
this result even when the procedure has been repeated {\it more
than $N$ times} and the (classical) fluctuations in the number of
atoms on the left and right sides exceed the natural (quantum)
fluctuations, of order $\sqrt{N}$, that were present in the
original BEC. In practice, the process cannot be repeated
infinitely many times, simply because the model of non-interacting
particles would break down at some point. However, this limitation
has nothing to do with the effect that we are considering, and the
system can be designed such that it only appears after a number of
repetitions that is much larger than $N$.

It is worth mentioning two remarks regarding the procedure
involving the stream of flying atoms. (1) When we traced over the
target atoms' degrees of freedom, we have excluded knowledge about
the BEC density matrix that we would have obtained by keeping a
record of the outcome of measurements on those atoms. However,
since we are looking for correlations between the states of the
target atoms only, and we are not interested in correlations
between those and the BEC, it is safe to trace over their degrees
of freedom when considering subsequent repetitions of the
procedure. (2) It is natural to think of the BEC as a catalyst for
converting the entanglement, rather than contributing its own
entanglement to the resulting pairs of target atoms. That can be
seen by observing the fact that the BEC can be reused in the
procedure indefinitely.

\section{Conclusion}

We have discussed the fundamental difficulty with directly
observing nonlocal quantum effects, e.g. testing Bell
inequalities, using the entanglement between spatially separated
modes of massive particles. These difficulties apply to some basic
procedures attempting to convert this mode entanglement into
Bell-testable forms. We have then analyzed several possible
approaches where this conversion can be achieved by using
auxiliary particles, including the case of an auxiliary BEC.
Remarkably we found that in the case of a BEC, the number of times
we can convert such entanglement is {\it not} limited by the
number of particles in the BEC. Although we have not given
mathematical proofs, we suspect that the procedures we propose
provide the maximum obtainable conversion probabilities using a
given number of particles. Our results using a single auxiliary
atom apply regardless of whether the particles are bosons or
fermions. However, we cannot see any simple analogue of the BEC
method in the case of fermionic flying atoms.

\begin{acknowledgments}
We would like to thank T. Iitaka, A. Lvovsky, D. Markham and A.
Zavatta for useful discussions. This work was supported in part by
the National Security Agency (NSA), the Laboratory for Physical
Sciences (LPS) and the Army Research Office (ARO); and also by the
National Science Foundation (NSF) grant No.~EIA-0130383. One of us
(S.A.) was supported by the Japan Society for the Promotion of
Science (JSPS).
\end{acknowledgments}


\begin{thebibliography}{99}
\bibitem{NonlocalityWarning} The exchange of classical information
required in order to establish the presence of quantum
correlations is not included here. That step will naturally take
longer than $d/c$.
\bibitem{Measuring_modes} Here we have in mind local measurements. We
therefore do not include the possibility of measuring
superpositions of different distributions of a well-defined number
of particles between multiple, spatially separated modes
\cite{Leggett}.
\bibitem{MeasurementNote} In actual experiments, the violation of
Bell inequalities using single photons was observed using direct
measurements on the photons. The methods used there require
knowledge of quantum-optics techniques. We shall therefore not
explain them here \cite{Babichev,DAngelo}.
\bibitem{Experiment} For concrete possible implementations of this
process and others discussed in this paper, see e.g. Refs.
\cite{Zhang,Calarco,Raizen}. We stress, though, that we are mainly
interested in fundamental questions here, and one can think of the
experiments we propose here as thought experiments. We only
mention certain experimental procedures in order to demonstrate
that there is no fundamental objection to the processes we
discuss.
\bibitem{Merging_BEC} Here we need that a target atom in the ground
state favours zero occupation of the flying atom trap, whereas a
target atom in the excited state favours single-particle
occupation of the trap. This situation could in principle be
achieved using a combination of target-atom-mediated interactions
between the flying-atom species and trapping-well-dependent
interaction resonances.
\end{thebibliography}
\end{document}